\documentstyle[aps,preprint,tighten]{revtex}
\begin{document}
\draft
\title{Proximity Effect and Charging in Mesoscopic\\
Normal Metal - Superconductor Junction Systems}
\author{C. Bruder$^{a)}$, Rosario Fazio$^{b)}$, and Gerd Sch\"{o}n$^{a)}$}
\address{ a)Institut f\"ur Theoretische Festk\"orperphysik,
Universit\"at Karlsruhe, 76128 Karlsruhe, FRG\\
b)Istituto di Fisica, Universit\`a di Catania,
viale A. Doria 6, 95129 Catania, Italy\\}
\date{\today}
\maketitle
\begin{abstract}
Charging effects in mesoscopic junctions suppress the tunneling of electrons.
In normal metal - superconductor systems they also suppress the
proximity effect, thereby revealing the nature of
the microscopic processes and of the ground state of the system. The effect
can be made visible since the charging and proximity effect and hence
the transport properties of the system can be modulated by gate voltages. If
the superconducting electrode is isolated, electron number parity effect lead
to an even-odd asymmetry in the proximity effect.

\end{abstract}

Single electron charging effects influence the transport
through small capacitance junction systems \cite{singlecharge}. If part of
the system is superconducting
new phenomena have been observed. They include parity effects in systems
containing superconducting islands \cite{harvard,saclay}, which arise since a
single unpaired electron in a superconducting island raises the energy by the
gap energy $\Delta$ \cite{avnaz}. This makes the correlated tunneling of
two electrons by the process of an Andreev reflection an important channel
for charge transfer \cite{hekglaz}. Charging effects also
influence the supercurrent through SSS or SNS structures with a mesoscopic
normal island in the middle \cite{matveev,bauern}.

When a superconductor and a normal metal are put in contact, Cooper pairs
leak to the normal region and a superconducting order
parameter is induced in the metal. Since charging effects inhibit
tunneling they reduce the proximity effect. Both can be
modulated by gate voltages, which opens the way to modulate supercurrents.
This creates the possibility to detect the effect and
interesting perspectives for future applications \cite{esteve}.

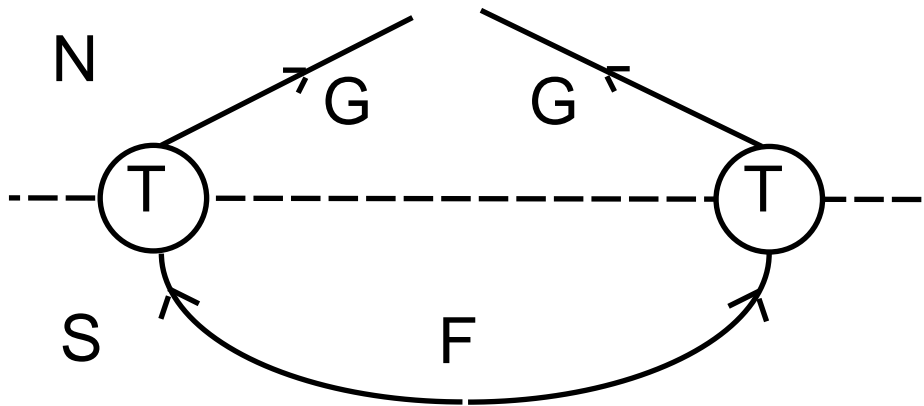
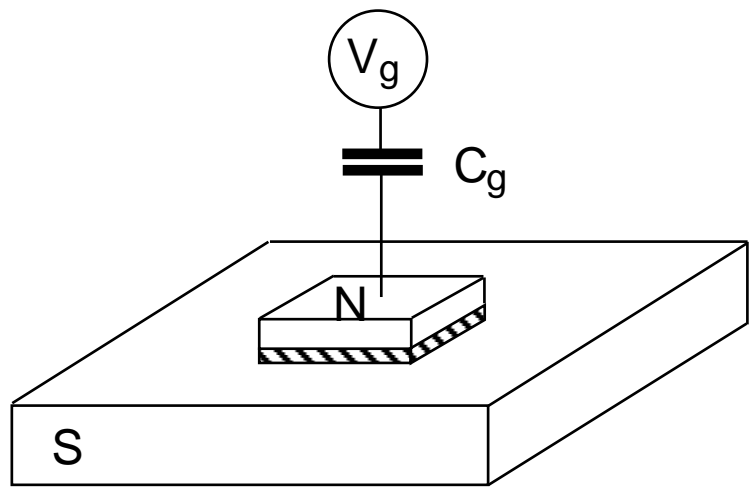
\begin{figure}
\caption{
Temperature dependence of the induced pair amplitude (in units of $Ge^2/hV$)
at $Q_g=0$ for $E_C/\Delta=0$ (upper curve), $E_C/\Delta=0.23$
(middle curve), and $E_C/\Delta=0.45$ (lower curve).
Upper inset:
One of the setups studied. A small normal island (N) is coupled
 to a superconductor (S) by a tunnel junction. The gate voltage
$V_g$ applied via a capacitor $C_g$ is  used to modulate the charging
effects.
Lower inset:
The induced pair amplitude in the normal metal (N) to second order in
the tunneling in diagrammatic language. F is the Gorkov Greens function
in the superconductor (S), G is the ordinary Greens function.
}
\label{fig3}
\end{figure}

The proximity effect had been studied in macroscopic junction systems by
Aslamazov, Larkin and Ovchinnikov (ALO)  \cite{ALO}. We show how
charging effects in mesoscopic systems modify their results.
We first study the system shown schematically in the upper inset of Fig.
\ref{fig3}: a small
normal island separated from a bulk superconductor (with gap $\Delta$)
by a low capacitance tunnel junction. The island is coupled capacitively to
a gate voltage. In the second part of this article we consider the situation
where the island is superconducting and study the proximity effect in the
normal electrode and the modification of its transport properties. In this
case electron number parity effects can be observed in the proximity
effect.

The system is described by the Hamiltonian $H = H_{0}+H_{t}+H_{ch}$
where $H_{0}$ refers to the electrons in the normal island
and the superconductor, $H_{t}$ describes tunneling, and $H_{ch}$
the capacitive Coulomb interaction. In the presence of the gate voltage source
the electrostatic charging energy is
$E_{ch}(q,Q_g) = (qe-Q_g)^2/2C$. Here $q$ is the number of excess electrons
on the island, $C$ its total capacitance, i.e. the sum of the junction
capacitance and that to the gate voltage. The gate voltage is
responsible for the offset charge $Q_g = q_g e = C_gV_g$.

The essence of the proximity effect is a nonvanishing pair amplitude
$\langle\psi_N \psi_N \rangle$ induced in the normal metal
by the coupling to the superconductor. The size of the induced
energy gap $\Delta_{ind} = \lambda \langle\psi_N \psi_N \rangle$ depends
further on the effective pairing interaction $\lambda$ in the
metal. To lowest order in the tunneling matrix element $\mid T \mid ^2$,
the pair amplitude is represented by the diagram shown in the lower inset of
Fig. \ref{fig3}.
It has been analyzed by (ALO)  \cite{ALO}. In order to evaluate the influence
of the charging energy (which is a nonperturbative effect) we formulate the
problem in an effective action approach, generalizing the work of
Ref. \cite{AES}. The partition function of the system can be recast as a
path integral over a  phase variable which is related to the voltage drop at
the junction interface $\phi(\tau) = (2e/\hbar) \int_0^\tau d\tau' V(\tau')$.
Since the total number of charges on the island is an integer, values of the
phase which differ by $4\pi$ have to be identified. Accordingly,
the partition function includes a summation over winding numbers
$\phi(\beta)=\phi(0)+4\pi n$ \cite{SZ}. In the second arrangement
with a superconducting island parity effects at low temperatures suppress
odd number charge states. A restriction to even number states corresponds
to the boundary conditions $\phi(\beta)=\phi(0)+2\pi n$. The charging
energy, including the effect of the gate voltage, is described by the action
\begin{equation}
	S_{ch}=\int_0^\beta d\tau \left[ \frac{C}{2}
	\left(\frac{\hbar}{2e}
	\frac{\partial \phi}{\partial \tau}\right)^2 +
	i Q_g \frac{\hbar}{2e} \frac{\partial \phi}{\partial \tau} \right]
\label{S_ch}
\end{equation}
The effect of tunneling, in an expansion in the tunneling matrix element, is
described by the action
\begin{eqnarray}
S_{t}=
\int_0^\beta d\tau \int_0^\beta d\tau' |T|^2 \left\{
G_N(\tau-\tau')G_S(\tau'-\tau)
\cos\left[\frac{\phi(\tau)-\phi(\tau')}{2}\right]\right .
\nonumber\\
\left . +F_N(\tau-\tau')F_S(\tau'-\tau)
\cos\left[\frac{\phi(\tau)+\phi(\tau ')}{2}\right]\right\}
\label{eq:Sqp}
\end{eqnarray}
Here $G_{S,N}(\tau)$ and $F_{S,N}(\tau)$ are quasiclassical diagonal and
off-diagonal Greens functions of the superconducting and normal electrodes,
respectively. The first contribution describes quasiparticle tunneling. At
low temperatures in an ideal system it is suppressed below the gap voltage,
but it sets in when the gain in charging energy exceeds the energy gap for
quasiparticle excitations. The second term (written here for later comparison)
vanishes if one of the electrodes is normal, $F_N = 0$. If both
electrodes are superconducting it describes Cooper pair tunneling
in the general form (incl. the so-called "quasiparticle - Cooper pair
interference" effect).
The two electrons tunnel in a correlated way at times $\tau$ and $\tau'$
within the time scale $\Delta^{-1}$. Higher order terms proportional to
$\mid T \mid ^4$ describe further correlated two-particle tunneling.
This includes Andreev reflection which becomes an important channel for
charge transfer, when the gap $\Delta$ exceeds the scale of the charging
energy $E_C = e^2/2C$.

The pair amplitude $\langle\psi_N \psi_N \rangle$ in the normal metal can be
calculated by adding a source term to the Hamiltonian
$H_{\lambda}= \int_0^{\beta} d\tau \int d^3 r \lambda(\bbox{ r},\tau)
\psi_N(\bbox{ r},\tau)\psi_N(\bbox{ r},\tau)$
and taking the functional derivative
$
\langle\psi_N \psi_N \rangle_{\bbox{ r},\tau_0} = -  \left . \delta \ln (Z)/
\delta \lambda(\bbox{ r},\tau_0) \right | _{\lambda=0}
$.
This produces a term similar to the second one in (\ref{eq:Sqp}),
generalized to the situation where the magnitude of the order parameter
depends on space and time. Noting that the Gorkov Greens function $F_S$ in
the superconductor is short-ranged in space and the tunneling matrix elements
are localized to the junction interface ($\bbox{ r}=\bbox{ \rho}$)
we can write the pair amplitude as
\begin{equation}
\langle\psi_N \psi_N \rangle_{\bbox{r_0},\tau_0}={G \over G_0}{V
\over N_N(0)S} \int d\tau \int d\tau'\int d^2\rho
K(\bbox{ r_0},\bbox{ \rho},\tau,\tau';\tau_0)
F_S(\bbox{ \rho},\bbox{ \rho},\tau-\tau')
h(\tau,\tau';\tau_0)
\label{eq:psipsi'}
\end{equation}
Here $G$ is the
conductance of the junction and $G_0=4e^2/h$ the quantum conductance.
The kernel $K$ is related to the Cooperon
propagator $\bar{G(\bbox{ r}-\bbox{ \rho},\tau-\tau_0) G^*(\bbox{ r}-\bbox{
\rho},\tau'-\tau_0)}$. Its Fourier transform can be expressed as
\begin{equation}
K(\bbox{r},\bbox{\rho},\omega_\nu,\omega_\eta) = N_N(0) \int d\xi \int
d\xi'
\frac{ p(\bbox{r},\bbox{\rho},\xi - \xi')}{(\xi - i\omega_\nu )
(\xi' + i\omega_\eta )}
\end{equation}
The function $p(\bbox{r},\bbox{\rho},t)$ satisfies a diffusion equation with
the boundary condition that the probability current normal to the surface
of the island vanishes \cite{deGennes}.

In the classical case the function $h=1$, and the ALO result is recovered.
The quantum fluctuations due to charging are taken into account by the
phase correlator
$
h(\tau,\tau';\tau_0)= \langle \exp\{i
(\phi(\tau)+\phi(\tau')-2\phi(\tau_0))/2\}\rangle_{S_{ch}} .
$
The expectation value has to be calculated using the charging action
(\ref{S_ch}) including a summation over winding numbers (note that the
weight for $Q_g \ne 0$ is complex). The function $h(\tau,\tau';\tau_0)$ is
the amplitude for forward single electron transfers  at times
$\tau$ and $\tau'$ (similar to the term involving
$\cos\left[\frac{\phi(\tau)+\phi(\tau')}{2}\right]$ in $S_t$ )
and a 2e transfer back at time $\tau_0$. The correlator, evaluated
in the charge representation, is
\begin{equation}
\langle \exp\left[i\frac{\phi(\tau)+\phi(\tau')}{2} -
i\phi(\tau_0)\right]\rangle_{S_{ch}}
 =  \exp(E_C |\tau'-\tau|)
\langle \exp{\{2E_C[1+(q-q_g)](\tau+\tau' - 2\tau_0))\}}\rangle_q
\label{eq:cos}
\end{equation}
for $\tau < \tau' < \tau_0$ and similar expressions for other relations
between $\tau, \tau'$ and $\tau_0$.
The average $\langle \rangle_q$ is taken by summing over discrete charge
states $q = 0, \pm 1,
\pm 2, ....$ with weight $\exp{[-\beta E_C(q-q_g)^2]}$ properly normalized.

The charging effects are most important in the case of a very small island
$\hbar D/d^2 \gg \max(E_C,\Delta )$ (here d is the linear dimension of the
island). In this case the propagator
$p(\bbox{r_1},\bbox{r_2},\xi - \xi') \sim  V^{-2} \delta(\xi - \xi')$
is essentially a constant; substituting this expression in eq.
(\ref{eq:psipsi'}) we obtain
\begin{equation}
\langle\psi_N \psi_N \rangle = \frac{ G V}{G_0\beta^3}
\sum_{\omega_\mu} \sum_{\omega_\nu,\omega_\eta >0}
\frac{\Delta}{(\omega_\nu+\omega_\eta) \sqrt{\omega_\mu^2 + \Delta^2}}
h(\omega_\nu - \omega_\mu ,\omega_\mu - \omega_\eta)e^{i\chi}\;.
\label{eq:psipsi'final}
\end{equation}

In larger islands
$p(\bbox{r_1},\bbox{r_2},t)$ describes the decay of the pair amplitude.
In the absence of inelastic and pairbreaking effects, the decay length is
given by the thermal diffusion length $\sqrt{\hbar D/T}$.
Pair breaking effects can be easily incorporated. They
influence the Cooperon propagator, and thus
proximity effect and decay length.

The modification of the pair amplitude as compared to the classical case
depends on temperature and applied gate voltage. The temperature dependence is
shown in Fig. \ref{fig3}. For $E_C/\Delta=0$ (upper curve) the pair amplitude
diverges logarithmically for $T\to 0$. This is the result of ALO \cite{ALO}.
For finite values of the charging energy the proximity effect
is suppressed; also the divergence at $T=0$ is removed. If the pairing
interaction $\lambda$ is finite, eq. \ref {eq:psipsi'final} acquires an
additional factor $(1+{\lambda m p_F \over \pi} T \sum _{\omega_\nu >0}
{1\over \omega_\nu})^{-1}$. For repulsive $\lambda$ this leads to an
more pronounced suppression of the induced pair amplitude at very low
temperatures and a reentrant behavior.

The charging effects and the proximity effect can be modulated by gate
voltages as shown in Fig. \ref{fig4}.
The periodicity of the modulation is $1e$, reflecting the
effect of single electron tunneling. Even though the rate of single electron
tunneling may be low for low voltages, it still is sufficient at low sweeping
rates to reset the state of the system in a setup where the island is
a normal metal.

In the second arrangement, where a small superconducting island is coupled to
a film of normal metal the modulation is even more pronounced.
In this case a single excitation
created in the superconductor does not find a partner to recombine. Hence
the energy of odd electron number states lies higher by $\Delta$ than that
of the corresponding normal island. As a result a single electron tunnels
at low temperatures only if the gain in charging energy exceeds the
cost in excitation energy. The excitation energy can be regained when
another electron enters the island
\cite{avnaz,SZparity}. This leads to the 2-e periodic long-short cyclic
dependence observed in the properties of SN electron boxes \cite{saclay} or
transistors \cite{harvard} and studied theoretically in Refs.
\cite{matveev,bauern}. As long as the temperature is low we can
describe the new situation by summing in (\ref{eq:cos}) only over even charges
$q = 0, \pm2, ...$ in the range where even electron number states have
the lower energy, $E_{ch}(\text{even},Q_g) < E_{ch}(\text{even} \pm1,Q_g)
+\Delta$, or only
over odd charges $q = \pm 1, \pm 3, ...$ in the range where odd states
win. The pronounced dependence of the induced pair
amplitude as a function of the gate voltage is shown in
Fig. \ref{fig5}. At low temperatures the pair amplitude diverges
logarithmically at the points of degeneracy $E_{ch}(\text{even},Q_g) =
E_{ch}(\text{even}
\pm1,Q_g) +\Delta$. Only at these points the low temperature
divergence of the ALO result is not removed.
In either case the divergence is an artifact of the perturbation theory. It
is regularized by higher order tunneling processes, which give rise to a
lifetime broadening \cite{AZ}.

In the case where  $\Delta > E_C$ quasiparticle tunneling remains
suppressed for all values of the gate voltage. In this case the process which
transfers charges and resets the system is Andreev reflection
\cite{hekglaz,GS}. A 2e-periodic picture emerges with peaks in the
induced pair amplitude at odd integer values of $q_g$ (see
Fig. \ref{fig5}).

The modulation of charging effects helps to make the proximity effect
visible. As an application we consider the situation where a
superconducting electrode is placed on top of a normal film.
If the film is thin enough a uniform  pair amplitude will be induced below
the electrode.
A current through this normal film depends on the induced superconductivity
\cite{esteve,pannetier}. If the electrodes have a low capacitance and are
coupled
to the voltage source, the current through the normal film can be modulated.

In summary we have presented an effective action description of charging
effects in normal metal-superconductor tunnel junctions and shown that the
Coulomb interaction in the island suppresses the proximity effect.
We recover the classical results of ALO. The charging is  accounted for by an
extra phase correlation function modifying the classical expressions. It
removes the low temperature divergences obtained in the
classical limit. Several extensions can be included, e.g. relaxation processes
due to the flow of Ohmic currents and the effect
of a gap in the normal island induced by the proximity effect. Both
regularize the divergence of the classical result.

\acknowledgements
We thank D. Esteve for stimulating discussions initiating this work, and
 A.I. Larkin,
B. Pannetier, A. van Otterlo, F.~W.~J. Hekking, Yu.~V. Nazarov, G. Falci,
and A.~D. Zaikin for many useful comments.
We acknowledge the hospitality of the ISI, Torino, and the support of
the 'Sonderforschungsbereich 195' of the DFG.

\begin{figure}
\caption{
Dependence of the proximity effect on the induced charge $Q_g = C_g V_g$ for
$T/T_C=0.05$, $E_C/\Delta=0.23$ (upper curve)
$E_C/\Delta=0.45$ (middle curve) and $E_C/\Delta=0.7$ (lower curve).
The curves are periodic in $Q_g$ with period $1e$.
}
\label{fig4}
\end{figure}

\begin{figure}
\caption{
Dependence of the pair amplitude on  $Q_g$
in the case where the island is superconducting and parity effects reduce
the regime of the odd electron number state for $T=0$
and (a)  $E_C/ \Delta=0.45 < 1$, for curve (b) $E_C/\Delta=1.8 >1$
(no single electron transitions). The curves have now period $2e$.
}
\label{fig5}
\end{figure}

\end{document}